\documentclass[reprint,           
               twocolumn,
               noshowpacs,          
               nopreprintnumbers,     
               aps,                 
               prl,                 
               letter,             
	       superscriptaddress,  
               nofootinbib,         
               tightenlines,        
               floats,floatfix    
               ]{revtex4}

\usepackage{amsmath, amsfonts, amsthm, amssymb, graphicx}
\usepackage{color}
\usepackage[utf8]{inputenc}
\usepackage{lipsum}
\usepackage{centernot}
\usepackage{ulem}
\def\be{\begin{equation}}
\def\ee{\end{equation}}
\def\ba{\begin{eqnarray}}
\def\ea{\end{eqnarray}}
\def\beastar{\begin{eqnarray*}}
\def\eeastar{\end{eqnarray*}}       

\def\bdm{\begin{displaymath}}
\def\edm{\end{displaymath}}

\def\bq{\begin{quote}}
\def\eq{\end{quote}}

 at 10truept

\newcommand{\beq}{\begin{equation}}
\newcommand{\eeq}{\end{equation}}
\newcommand{\bea}{\begin{eqnarray}}
\newcommand{\eea}{\end{eqnarray}}
\newcommand{\beqa}{\begin{eqnarray}}
\newcommand{\eeqa}{\end{eqnarray}}

\def\bnabla{{\bar \nabla}}
\def\bBox{{\bar \Box}}
\def\bg{{\bar g}}

\newcommand{\rd}{\mathrm{d}}

\newcommand{\citeseq}{\cite{KP1,KP2,KP3,KPSZ, KPS, KP4,etude}}
\def\ltap{\ \raise.3ex\hbox{$<$\kern-.75em\lower1ex\hbox{$\sim$}}\ }
\def\gtap{\ \raise.3ex\hbox{$>$\kern-.75em\lower1ex\hbox{$\sim$}}\ }
\def\gl{\ \raise.5ex\hbox{$>$}\kern-.8em\lower.5ex\hbox{$<$}\ }
\def\roughly#1{\raise.3ex\hbox{$#1$\kern-.75em\lower1ex\hbox{$\sim$}}}

\begin{document}

\title{Gravitational Mechanisms to Self-Tune the Cosmological Constant: Obstructions and Ways Forward}

\author{Florian Niedermann} 
\email{florian.niedermann@nottingham.ac.uk}
\author{Antonio Padilla} 
\email{antonio.padilla@nottingham.ac.uk}
\affiliation{School of Physics and Astronomy, 
University of Nottingham, Nottingham NG7 2RD, UK}

\date{\today}

\begin{abstract}
  Gravitational models of self-tuning are those in which vacuum energy
  has no observable effect on spacetime curvature, even though it is a
  priori unsuppressed below the cut-off. We complement Weinberg's
  no-go theorem by studying field-theoretic completions of
  self-adjustment allowing for broken translations {as well as other
    generalisations}, and identify new obstructions. Our analysis uses
  a very general K\"all\'en-Lehmann spectral representation of the
  exchange amplitude for conserved sources of energy-momentum and
  exploits unitarity and Lorentz invariance to show that a transition
  from self-tuning of long wavelength sources to near General
  Relativity on shorter scales is generically not possible. We search
  for novel ways around our obstructions and highlight two interesting
  possibilities. The first is an example of a unitary field
  configuration on anti-de Sitter space with the desired transition
  from self-tuning to GR. A second example is motivated by vacuum
  energy sequestering.
\end{abstract}
\maketitle

\section{Introduction}

The cosmological constant problem \cite{zeldovich, wein, cliff, pol,
  me} follows automatically from our ``best" model of Nature at low
energies, in which matter is described by a local quantum field theory
(QFT) minimally coupled to a gravitational sector described by
classical General Relativity (GR). In the absence of a suitable
symmetry mechanism, virtual particles endow the vacuum with an energy
density scaling like the fourth power of the effective field theory
cut-off.  In GR, this vacuum energy gravitates like a cosmological
constant, curving the spacetime geometry even in vacuum. Cosmological
observations constrain the net cosmological constant to be no greater
than the dark energy scale, $(\textrm{meV})^4$, at least sixty orders
of magnitude below the theoretical estimate based on a TeV-scale
cut-off, beyond which new symmetries may emerge. The precise details
of this tuning are extremely sensitive to the small print of
high-energy physics, in complete violation of the naturalness criteria
\cite{chiral}. Because naturalness plays such an important role in
many aspects of particle physics \cite{nat}, its most notable failure,
the cosmological constant problem, is generally regarded as one of the
most important problems in Theoretical Physics.

The scarcity of viable proposals for solving the cosmological constant
problem has led some to abandon naturalness, and seek anthropic
explanations within a scannable landscape of vacua \cite{pol,Bousso}.
Nevertheless, there do exist natural alternatives to this, most
notably {\it vacuum energy sequestering} \citeseq (more on that
later).  In this letter we are interested in so-called {\it
  self-tuning}, or {\it self-adjusting} proposals, in which the vacuum
energy is a priori unsuppressed below the cut-off, but has no
significant observable effect on the spacetime curvature thanks to the
adjustment of new fields.  Such scenarios are famously forbidden by
Weinberg's no go theorem \cite{wein}, at least if we assume a
local kinetic sector and translational invariance of the vacuum
solution. If the latter assumption is relaxed, self-tuning can be
possible, as in the so-called {\it Fab Four} scenario
\cite{Fab1,Fab2,Fab3}, although the presence of a light scalar field
in the gravitational interaction poses potential problems for short
distance phenomenology.

Self-tuning is also realised in the 5D braneworld set-up
of \cite{selftun}, at the price of introducing a singular bulk. This
singularity is tamed in a recent proposal \cite{cc}, although it
remains to be seen whether it yields a healthy perturbative
description with viable phenomenology. {Moreover, much of the
  past effort had its focus on 6D braneworld models where the vacuum
  energy curves the bulk into a cone. However, whilst the finite
  volume proposal~\cite{Sled0,Sled} relies on a radiatively
  unprotected parameter tuning ~\cite{noSled1,noSled2}, the infinite
  volume realisation~\cite{BIG0,BIG_K} either suffers from a ghost
  instability or lacks a 4D gravity
  regime~\cite{BIG-1,BIG-2,BIG1,BIG2}. In fact, the pathology
  encountered in the latter case will fall in the class of new
  obstructions discussed here.}

The purpose of this work is to identify new obstructions to
self-tuning that complement Weinberg's approach and look for some
novel ways in which we might overcome them. By self-tuning, here we
really mean the absence of a geometrical response in spacetime to
sources of infinite wavelength, even when localised sources gravitate
normally, as required by short distance gravity tests. By this
definition, we do not consider the possibility of self-tuning by scale
invariance, in which the ratio of the field theory masses and the
scale of gravity runs to zero, already captured by Weinberg's analysis
\cite{wein}, and incompatible with our low energy
Universe, owing to the observed particle masses relative to the Planck
scale.

Our approach to the self-tuning question is one which incorporates a
complete field theoretic description, allowing for configurations that
break translation invariance {and have non-local kinetic
  operator},\footnote{{The only requirement is that the kinetic
    operator admits a K\"all\'en-Lehmann spectral decomposition. The
    braneworld inspired propagator discussed in \cite{non-local2}
    provides a prototypical example (see also \cite{non-local1}).}} in
constrast to Weinberg, and factoring in constraints from observational
tests of GR up to solar system scales~\cite{test}. To this end, we
consider the exchange amplitude for two conserved sources of energy
momentum in a background Minkowski spacetime, mediated by both single
and multi-particle states up to spin-2. We use a K\"all\'en-Lehmann
(KL) spectral representation \cite{Kallen,Lehmann} to capture the
fully quantum corrected exchange amplitude that is linear in each
source. We assume that the intermediate states couple linearly to
sources and that the free field propagators are of their canonical
form, compatible with unitarity and Lorentz invariance. By further
imposing unitarity through the positivity of the spectral density we
are able to show that generically {\it one cannot pass from a
  self-tuning regime for sources of infinite wavelength to one in
  which we recover GR to sufficient precision for localised sources. }
We then look for novel ways in which we can get around our
obstructions.

Our analysis can be extended to a background de Sitter spacetime with
the same generic results but not to a background anti-de
Sitter. Indeed, one can find explicit examples of linearised field
configurations in anti-de Sitter that self-tune at large distances but
recover GR on shorter scales with the correct tensor structure.  We
identify a second way around our obstructions motivated by vacuum
energy sequestering \citeseq, being of greater phenomenological
interest. There we find that the structure of the free field
propagator is not always canonical, but exhibits features reminiscent
of the decapitation scenario \cite{decap}.  A third possible way
around involves so-called screening mechanisms
\cite{cham1,cham2,Vain1,Vain2,Vain3,symm1,symm2} which rely on
non-linear couplings to the source.


\section{Obstructions to self-tuning in Minkowski spacetime}
Consider the exchange amplitude for two conserved sources,
$ T_{\mu\nu} $ and $ T'_{\mu\nu} $, on a Minkowski background, mediated by
both single- and multi-particle states up to spin-2. Since we assume
{\it linear} couplings to {\it conserved} sources, spin-1 states
cannot couple to the source, so we only need to consider intermediate
states of spin-0 and 2.  Assuming Lorentz invariance and unitarity, we
can express the amplitude, using a spectral representation, as
\begin{multline}
  {\cal A}=\bar \rho_2 \, \bar {\cal A}_2+\bar \rho_0 \, \bar {\cal A}_0 + \\
  \int_{0^+}^\infty \rd s \, \rho_2(s) \, {\cal A}_2(s) + \int_{0^+}^\infty \rd s \, \rho_0(s) \,  {\cal A}_0(s) \label{amp} \;,
\end{multline}
where we have included the exchange of a single massless spin-2 state
with coupling $\bar \rho_2$, massive spin-2 states of mass $s$ and
spectral density $\rho_2(s)$, a massless spin-0 state with coupling
$\bar \rho_0$, as well as a spin-0 states of mass $s$ and spectral
density $\rho_0(s)$. {The form of the free field propagators are those
  of canonical fields with the appropriate mass and spin, fixed by
  Lorentz invariance and unitarity. The corresponding one-particle
  exchange amplitudes} are specified in the Appendix in
Eqs.~\eqref{eq:A0gen} and \eqref{Amgen} for spin-2 (after setting
$\kappa=0$), whereas for spin-0 we have
\begin{multline} \label{eq:spin0}
  \bar {\cal A}_0 =
  \int \rd^4 x \, \sqrt{-\bar{g}} \; T' \frac{1}{-\bBox} \, T\;, \\
  \quad {\cal A}_0(s)=\int \rd^4 x \, \sqrt{-\bar{g}} \; T'
\frac{1}{-\bBox+s} \, T \;,
\end{multline}
where $\sqrt{-\bar{g}}=1$ in the Minkowski case.
The Green's function acting on a tensor of arbitrary rank is given by
$\left(\frac{1}{-\bBox+\mu^2} \, J^{\ldots}_{\ldots}\right)(x) = \int
\rd^4 y\, G_\mu(x, y) \, J^{\ldots}_{\ldots}(y)$, where
$(-\bBox_x+\mu^2) \, G_\mu(x, y)=\delta(x-y)$.  In general, the
precise form of these Green's functions will not be important for our
discussion, although we will state certain important properties where
appropriate.  Unitarity requires us to assume positivity of the
spectral densities, i.e.\
$\bar \rho_2,\, \bar \rho_0, \, \rho_2(s),\, \rho_0(s) \geq 0$.

If this theory is to ``self-tune" along the lines described in the
introduction, the amplitude should vanish in the presence of an
infinite wavelength vacuum energy source,
$T_{\mu\nu} = -V_{vac} \, \eta_{\mu\nu}$, and a localised probe,
$T'_{\mu\nu}$, or in other words
\begin{align}
  \frac23 \, \bar \rho_2 \, \frac{1}{-\bBox~} \, 1 - 4 \, \bar \rho_0 \, \frac{1}{-\bBox~} \, 1 \,
  - 4 \, \int_{0^+}^\infty \rd s \, \rho_{0}(s) \, \frac{1}{s} \,1 = 0 \;, \label{form}
\end{align}
where the convolution of the Green's function with unity is
understood, and we have used the fact that for $ s \neq 0$,
$\frac{1}{-\bBox+s} \, 1=\frac1s$.  Moreover, the massless Green's
function was assumed to be Lorentz invariant ensuring the vanishing of
the transverse-tracefree part of the energy-momentum tensor
$T_{\mu\nu}^{(\rm TT)}$ for a constant vacuum energy source (see
Appendix).
If we operate on Eq.\ \eqref{form} with $\bBox$, we obtain the
``self-tuning'' relation: $\bar \rho_2 = 6 \, \bar \rho_0$.

In contrast, for localised sources at shorter wavelength, we demand
close agreement with GR in order to recover its experimental success
\cite{test}, specifically
\begin{equation}
  {\cal A} \to \frac{1}{M_{\rm Pl}^2} \int \rd^4
x \, \left[ T'^{\mu\nu} \frac{1}{-\bBox~} \, T_{\mu\nu} - \frac12
  \left(1-\epsilon\right) T' \frac{1}{-\bBox~} \, T\right] \label{uv} \;,
\end{equation}
as $-\bBox \to \infty$, with $|\epsilon| \lesssim 10^{-5}$. We
  use Eqs.~\eqref{eq:spin0}, \eqref{A0} and \eqref{Am} to express
  $\mathcal{A}$ in terms of the spectral densities and sources. The
  above condition then imposes two ultra-violet constraints, which can
  be obtained by comparing the tensor ($\propto T^{\mu\nu}$) and
  scalar ($\propto T$) contributions to $\mathcal{A}$ independently. We express them in momentum space as
\begin{subequations}
  \label{cond}
   \begin{align}
    \label{cond1} 
  \frac{ \bar \rho_2}{x}+ \int_{0^+}^\infty \rd s \, \frac{\rho_2(s)}{x+s}   &\to \frac{1}{M_{\rm Pl}^2} \, \frac{1}{x} \;,
\end{align}
\begin{align}
  \label{cond2}
  \frac13 \, \int_{0^+}^\infty \!\!\rd s \, \frac{\rho_2(s)}{x+s} +2 \, \frac{\bar \rho_0}{x} + 2 \int_{0^+}^\infty \!\! \rd s \, \frac{\rho_0(s)}{x+s}   &\to \frac{\epsilon}{M_{\rm Pl}^2} \, \frac{1}{x} \;,
\end{align}
\end{subequations}
as $x \equiv p_\mu p^\mu \to \infty$.
Now, thanks to positivity of the spectral densities, we have that
$\frac{ \bar \rho_2}{x}$, $\frac{ \bar \rho_0}{x}$,
$\int_{0^+}^\infty \rd s \, \frac{\rho_2(s)}{x+s}$,
$\int_{0^+}^\infty \rd s \, \frac{\rho_0(s)}{x+s} \geq 0$, $\forall x \geq 0$,
and so from Eq.\ \eqref{cond2} we infer that $\bar \rho_0$,
$x \int_{0^+}^\infty \rd s \, \frac{\rho_2(s)}{x+s}$,
$x \int_{0^+}^\infty \rd s \,\frac{\rho_0(s)}{x+s} \lesssim |\epsilon
|/M_{\rm Pl}^2$ at large $x$. By the condition in Eq.\ \eqref{cond1},
we now find that $\bar \rho_2 \sim 1/M_{\rm Pl}^2$.  The ultra-violet
conditions $\bar \rho_2 \sim 1/M_{\rm Pl}^2$ and
$\bar \rho_0\lesssim |\epsilon |/M_{\rm Pl}^2$ are then in
contradiction with the "self-tuning" condition,
$\bar \rho_2 = 6 \, \bar \rho_0$, obtained in the infra-red - we do
not seem able to self-tune at large wavelength and recover GR at short
wavelength. Modulo our initial assumptions, this represents a
significant obstruction to self-tuning on a Minkowski background.

\section{Obstructions to self-tuning in de Sitter and anti-de Sitter spacetime}
We shall now attempt to extend our analysis to a background de Sitter
or anti-de Sitter spacetime, with curvature $\kappa$.  Of course, we
normally associate the notion of self-tuning with Minkowski vacua,
although we can easily extend its notion to other maximally symmetric
vacua by requiring that vacuum energy sources do not
gravitate. The so-called Fab 5 models are an example of this in a de
Sitter background \cite{fab5}. In any event, the exchange amplitude
for conserved sources, analogous to Eq.\ \eqref{amp}, is given by
\begin{multline}
  {\cal A}=\bar \rho_2 \, \bar {\cal A}_2+\bar \rho_0 \, \bar {\cal A}_0 + \\
  \int_{2 \kappa^+}^\infty \rd s \, \rho_2(s) \, {\cal A}_2(s) + \int_{s_0}^\infty \rd s \, \rho_0(s) \,  {\cal A}_0(s) \;,
\end{multline}
with the respective ($\kappa$-dependent) amplitudes defined in the
Appendix and Eq.\ \eqref{eq:spin0}. Although this amplitude is
the natural extension of the Minkowski one discussed before, we
hesitate to interpret it as a KL representation of the full quantum
corrected amplitude on account of the non-trivial asymptotic structure
(see \cite{DF,Bros,Ep} for a discussion of KL representations in de
Sitter and anti-de Sitter). The lower limits on the integrals are
fixed by the stability of the theory. In particular, perturbative
unitarity imposes the constraint on the spin-2 limit
$s \, > \, 2\kappa$ \cite{Higuchi,Deser, PM1, PM2, PM3}.  In anti-de
Sitter ($\kappa<0$), we impose the Breitenlohner-Freedman bound on the
scalar masses, $s_0 = 4 \kappa$ \cite{BF}{\footnote{{Since the range
      of integration for the massive states in AdS passes through
      zero, with massless states included explicitly elsewhere via
      $\bar \rho_0$ and $\bar \rho_2$, we shall set
      $\rho_0^{AdS}(0)=\rho_2^{AdS}(0)=0$. This technical point is
      included only in the interests of rigour -- it will play no role
      in our analysis.}}. We are not aware of an analogous bound in de
  Sitter ($\kappa>0$), so for now we conservatively impose $s_0 {>} 0$
  .
 
  Again, for ``self-tuning" we demand that the amplitude should vanish
  for a vacuum energy source, $T_{\mu\nu}=-V_{vac} \, \bg_{\mu\nu}$,
  that is
 \begin{multline}\label{condir2}
 \frac{\bar \rho_2}{6\, \kappa} +4 \, \bar \rho_0\, \frac{1}{- \bBox} \, 1 - \frac{1}{3}\, \int_{2 \kappa^+}^{\infty} \!\!\! \rd s \frac{\rho_2(s)}{ s - 2 \kappa } + 4 \int_{s_0}^\infty \!\!\! \rd s \, \frac{ \rho_0 (s)}{s} =0 \;.
 \end{multline}
Acting with $\bBox$ on this equation sets $\bar \rho_0 = 0 $.
In the ultra-violet, we again demand that Eq.\ \eqref{uv} holds,
implying a modified version of Eqs.~\eqref{cond} (after performing the
same steps as before),
\begin{subequations}
  \label{condds}
  \begin{align}
    \label{condds1} 
    \frac{ \bar \rho_2}{x}+ \int_{2 \kappa^+}^\infty \rd s \, \frac{\rho_2(s)}{x+s}   &\to \frac{1}{M_{\rm Pl}^2} \, \frac{1}{x} \;,
\end{align}
\begin{multline} \label{condds2}
  \frac13 \, \int_{2 \kappa^+}^\infty \!\!\rd s \, \frac{\rho_2(s)}{x+s} + \frac{2 \, \kappa}{3} \, \int_{2 \kappa^+}^\infty \!\!\rd s \, \frac{\rho_2(s)}{x+s} \, \frac{1}{s - 2 \kappa}  \\
+{2\frac{\bar \rho_0}{x} } + 2 \int_{s_0}^\infty \!\! \rd s \, \frac{\rho_0(s)}{x+s} \to
  \frac{\epsilon}{M_{\rm Pl}^2} \, \frac{1}{x} \;,
\end{multline}
\end{subequations}
as $x \to \infty$. For a de Sitter background ($\kappa>0$), at large
$x$, positivity of the spectral densities guarantees the positivity
of each individual term in Eq.\ \eqref{condds2}. It then follows
that
$x \int_{2 \kappa}^\infty \rd s \, \frac{\rho_2(s)}{x+s} \lesssim
|\epsilon |/M_{\rm Pl}^2$, enabling us to infer, using Eq.\
\eqref{condds1}, that $\bar \rho_2 \sim 1/M_{ \rm Pl}^2$.
Furthermore, as $x \to \infty$, we may also infer that
\begin{align}  \frac{|\epsilon |}{M_{\rm Pl}^2} &\underset{\phantom{positiviy\;\;}}{\gtrsim} \kappa \, \int_{2\kappa^+}^\infty \rd s \, \frac{ \rho_2(s)}{s-2\kappa} \, \frac{1}{1+s/x} \nonumber\\
&\underset{\text{positivity}\;\;}{\gtrsim}  \kappa \, \int_{2\kappa^+}^{s_*} \rd s \, \frac{ \rho_2(s)}{s-2\kappa} \, \frac{1}{1+s/x} \nonumber \\
&\underset{\quad  x \gg s_* \quad}{\sim} \kappa \,  \int_{2\kappa^+}^{s_*} \rd s \, \frac{ \rho_2(s)}{s-2\kappa} > 0 \label{ineq}
\end{align}
{where $s_*$ is some arbitrarily large, but finite, constant. We
emphasize that this only applies in de Sitter space - no such
inferences can be made on an anti-de Sitter background ($\kappa<0$).}

Staying with de Sitter space, let us consider the infra-red condition
in Eq.~\eqref{condir2}. {By plugging in
  $\bar \rho_2 \sim 1/M_{\rm Pl}^2$, $\bar \rho_0 = 0$ and using the
  inequality~\eqref{ineq}, we obtain\footnote{This is true as long as
    the spectral density $\rho_2(s)$ does not exhibit any isolated
    singular behaviour at infinity.}
\begin{align}
24 \, \kappa \, M_{\rm Pl}^2 \int^\infty_{s_0} \rd s \,
  \frac{\rho_0(s)}{s}  {\sim} - 1 + \mathcal{O}(\epsilon) \;,
\end{align}
which for $\kappa >0$ is in contradiction with the positivity of
$\rho_0 (s)$.} {This extends the results of the previous section to de Sitter space.}

\section{The anti-de Sitter loophole}
In the previous section, we saw how our conclusions could not
trivially be extended to anti-de Sitter space. This raises the
question: is there an anti-de Sitter loophole we can exploit?  It
turns out that there is. For an explicit example, consider the following two-parameter family of spectral densities on a
background anti-de Sitter space:
  $ \bar \rho_2=\frac{1}{M_{\rm Pl}^2} \left( 1 - \alpha_2 \right)$,
  $\rho_2(s)=\frac{\alpha_2}{M_{\rm Pl}^2}\, \delta(s-\mu_2)$,
  $\bar \rho_0 =0$,
  $\rho_0(s)=\frac{\alpha_0}{6 M_{\rm Pl}^2}\, \delta(s - \mu_0 )$,
  with couplings
\begin{align}
  \label{eq:couplings}
  \alpha_2= \frac{\mu_0}{\mu_2} \, \frac{\mu_2 - 2\kappa}{\mu_0 + 4 \kappa}\,, \quad  \alpha_0 = \frac{- \mu_0}{\mu_0 + 4 \kappa}\;.
\end{align}
This describes a massless graviton as well as a single massive
graviton and scalar of mass squared $\mu_2 \in (2 \kappa, - \mu_0/2]$
and $\mu_0 \in (0, -4\kappa)$, respectively, consistent with unitarity
\cite{Deser} and stability \cite{BF}. It is easy to check that this
fulfils both the ultra-violet \eqref{condds} and the infrared
constraint~\eqref{condir2}.

\section{The sequestering loophole}
In vacuum energy sequestering \citeseq, vacuum energy does not
gravitate yet the theory recovers GR for localised sources. It is
instructive to see how it gets past the obstructions described in this
letter. To this end, consider the effective gravitational equation of
motion in the original global version of the theory \cite{KP1,KP2}
 \begin{equation}
 M_{\rm Pl}^2 \, G_{\mu\nu}=T_{\mu\nu}-\frac14 \langle T \rangle \, g_{\mu\nu}\;,
 \end{equation}
 where the spacetime average of the trace of the energy momentum
 source is given by
 $\langle T \rangle = \int \rd^4 x \, \sqrt{-g} \, T / \int \rd^4x \,
 \sqrt{-g}$. Although this particular version\footnote{This condition
   can be relaxed in local formulations of the sequestering proposal
   \cite{KPSZ,KP4}} of sequestering can only be consistent with our
 universe if the spacetime volume is finite, we can study the dynamics
 of fluctuations in a locally Minkowski frame, approximating the
 volume integrals as integrals over a very large but finite section of
 Minkowski space. As shown explictly in \cite{KP4}, this represents an
 excellent approximation especially when we are close to the maximum
 scale factor in the cosmological evolution. In any event, if we
 proceed in this way the exchange amplitude is given by
 $ {\cal A}= \frac12 \int \rd^4 x \, h_{\mu\nu} \left(T'^{\mu\nu} -
   \frac14 \, \langle T' \rangle \eta_{\mu\nu} \right)$, and so we
 find that
\begin{multline}
  {\cal A} = {\cal A}_{GR} + \frac{1}{6 M_{\rm Pl}^2} \, \int \rd^4 x \,   T' \frac{1}{-\bBox~} \, T \\
 - \frac{1}{6M_{\rm Pl}^2} \,  \int \rd^4 x \, \rd^4 y \, T(x) \, \delta G(x, y) \, T(y)\;, \label{seqA}
\end{multline}
where ${\cal A}_{GR}= \frac{1}{M_{\rm Pl}^2}\bar {\cal A}_2$ is the
standard GR amplitude, and we recall that $ \frac{1}{-\bBox~} \, T $
denotes a convolution with the corresponding Green's function
$\int \rd^4 y \, G_0(x, y) \, T(y)$. From this we also introduce the
``decapitated'' Green's function \cite{decap},
$\delta G(x, y) = G_0(x,y) - \frac{1}{V} \int \rd^4 z \, G_0(x, z) - \frac{1}{V}
\int \rd^4 z \, G_0(z, y) + \frac{1}{V^2} \int \rd^4 z \,  \rd^4w \, G_0(z, w)$ and
$V=\int \rd^4z$.  This has the property that it vanishes for exactly
zero momentum, i.e., it vanishes when it is convoluted with a constant
$\int \rd^4 y \, \delta G(x, y) = \int \rd^4 x \, \delta G(x, y)=0$. Furthermore,
at non-zero momentum it behaves like the usual canonical Green's
function, $G_0(x, y)$. In other words, when it acts on a localised
excitation $\delta T = T-\langle T \rangle$, it gives
$\int \rd^4 y \, \delta G(x, y) \, \delta T(y)=\int \rd^4 y \, G_0(x, y) \, \delta
T(y)$.

What does all this amount to? The amplitude in Eq.\ \eqref{seqA}
describes the exchange of a massless particle of spin-2, a massless
particle of spin-0, and a massless decapitated particle of spin-0 that
happens to be ghostlike.  For localised sources, the latter two
contributions cancel out, yielding a well behaved GR like
amplitude. The cancellation follows along similar lines to the
cancellation of the brane bending mode and the conformal ghost in
Randall Sundrum gravity \cite{RS2, GT,Giddings}.  In contrast, for a
constant, vacuum energy source\footnote{Strictly speaking, our
  derivation of the amplitude in Eq.~\eqref{seqA} is only valid down to
  an infra-red cut-off set by the background curvature of our finite
  cosmology. Nevertheless it is interesting to extend the regime of
  validity of our result to see, in principle at least, how {the obstructions of the previous sections} might be evaded.}, the decapitated ghost gets
decoupled. This leaves the remaining spin-2 and spin-0 particles to
combine in such a way as to force the vanishing of the amplitude, as
required for self-tuning.

Clearly the amplitude \eqref{seqA} has an exotic field theory
content, in contrast to our generic case in Eq.~\eqref{amp}, which assumed
that the free field propagators took on their canonical form, and did
not include the possibility of decapitated propagators as one finds in
this approach to vacuum energy sequestering. A more thorough analysis
of linearised theory in the various versions of vacuum energy
sequestering is certainly warranted.

\section{Discussion}
In this letter we have shown how difficult it is to eliminate the
effect of the radiatively unstable vacuum energy contribution through
``self-tuning" in a phenomenologically viable theory of gravity. By
studying generic exchange amplitudes for conserved sources of energy
momentum we have identified significant obstructions to self-tuning in
the far infra-red and recovery of GR (to sufficient accuracy) in the
ultra-violet:

\textit{Self-tuning models that admit a standard spectral
    representation in terms of massless and massive spin-0 and spin-2
    states are generically incompatible with unitarity on both
    Minkowski and de Sitter backgrounds, provided the source coupling
    is linear.}

  For example, our analysis suggests that naive completions of, say,
  the old self-tuning brane model \cite{selftun}, or the ``filter''
  proposal \cite{degrav} are likely to face significant challenges
  from either phenomenology or unitarity.

  There are ways around our obstructions, however, and we identify a
  couple, each of which should provide a seed of future research. For
  each loophole linearised self-tuning can take place at low momentum
  even when we recover linearised GR at high momentum.

  \textit{From a phenomenological perspective, the loophole suggested
    by vacuum energy sequester holds greatest promise.  It is achieved
    by trading some canonical free-field propagators for decapitated
    ones, along the lines introduced in \cite{decap}. This connection
    of ideas merits further investigation and is, perhaps, the most
    important result to come from our work.}

  A second loophole occurs if the background curvature is anti-de
  Sitter space. Whether or not this is of interest phenomenologically
  remains to be seen, although we think it is unlikely. The presence
  of the anti-de Sitter loophole presents something of a puzzle from
  the point of view of any cosmological observer probing physics well
  within the cosmological horizon, decoupled from the details of the
  far infra-red. This conclusion is too quick. Although the probe is a
  short distance one, to explore the possibility of self-tuning one
  must ask how it interacts with a very long wavelength source. In
  this sense one is not considering scattering processes that are
  genuinely decoupled from the infra-red. The structure of the
  asymptotic vacuum matters with some field configurations
  catastrophically destabilising the vacuum in some cases, but not
  others, through ghost-like excitations.

  A third loophole concerns our assumption that the sources couple
  linearly to the states mediating the force. This cannot capture
  so-called screening mechanisms, such as Vainshtein
  \cite{Vain1,Vain2,Vain3}, chameleons \cite{cham1,cham2} or
  symmetrons \cite{symm1,symm2}, where non-linear couplings to matter
  play a vital role. It would be very interesting, although
  non-trivial, to try to incorporate these effects into our
  analysis. We are not aware of any self-tuning models that claim to
  exploit either chameleons or symmetrons in order to recover GR on
  the appropriate scale. The Vainshtein loophole, on the other hand,
  was also pointed out in Ref.~\cite{Dvali:2006su} in the special case
  of a spin-2 resonance theory with non-constant mass
  $m^2 \propto \Box^\alpha$ ($\alpha <1$) and later used to realize
  self-tuning in the ``degravitation'' proposal in
  Ref.~\cite{non-local2}, however without offering a fundamental
  theory. The general problem with Vainshtein screening, of course,
  regards its regime of validity of the effective theory and whether
  or not it can be trusted in the regime where screening is required
  \cite{AGS,dave,zhou}.

\begin{acknowledgments}
\vskip.5cm

{\bf Acknowledgments}: 
  We would like to thank Christos Charmoussis, Stefan Hofmann, Nemanja Kaloper, and
  Peter Millington for useful discussions.  F.N and A.P. are funded by
  an STFC Consolidated Grant. A.P is also funded by a Leverhulme Trust
  Research Project Grant.
\end{acknowledgments}

\section{Appendix: exchange of a single graviton in maximally symmetric space}

Consider the amplitude for the exchange of a single graviton between
two conserved sources, $ T_{\mu\nu} $ and $ T'_{\mu\nu} $, on a
maximally symmetric spacetime, with curvature $ \kappa $. In what
follows we shall allow for localised sources as well as non-localised
sources corresponding to constant vacuum energy.  In general, it is
given by
\begin{equation}
{\cal A}=\frac12 \, \int \rd^4 x \, \sqrt{-\bar g} \, h_{\mu\nu}
T'^{\mu\nu} \;, 
\end{equation}
where $\bar g_{\mu\nu}$ is the background metric, and $h_{\mu\nu}$ the
graviton fluctuation due to a source $T_{\mu\nu}$  satisfying
\begin{multline}
\bBox \left( h_{\mu\nu}-h \, \bar g_{\mu\nu} \right) - 2 \bnabla_{(\mu }\bnabla_{|\alpha|} \, h^\alpha_{\nu)} + \bnabla_\mu\bnabla_\nu \,h - 2\kappa \, h_{\mu\nu}
\\
+ \bar g_{\mu\nu} \left( \bnabla_\alpha \bnabla_\beta \, h^{\alpha\beta} - \kappa \, h \right) - m^2 \left(h_{\mu\nu} - h \, \bar g_{\mu\nu}\right)= - \frac{2}{M_{\rm Pl}^2} \, T_{\mu\nu} \;, \nonumber
\end{multline}
where indicies are raised/lowered with the background metric,
$\bnabla$ is the background covariant derivative, $ h= h^\mu_\mu$ and
$m^2$ is the graviton mass.  Decomposing the metric with respect to
representations of the background symmetry, we write
\begin{equation}
h_{\mu\nu}=h_{\mu\nu}^{\rm (TT)}+ 2 \, \bnabla_{(\mu} A_{\nu)}+ 2 \, \bnabla_\mu
\bnabla_\nu \, \phi + 2 \, \psi \, \bg_{\mu\nu}  \;,
\end{equation}
where $\bnabla$ is the background covariant derivative. The tensor is
transverse and tracefree: $\bnabla^\mu \, h_{\mu\nu}^{\rm (TT)}=0$ and
$\bg^{\mu\nu} h_{\mu\nu}^{\rm (TT)}=0$. The vector is divergence-free,
$\bnabla^\mu A_\mu=0$.

For a graviton with mass $m^2=0$, we can set $A_\mu=0,~\phi=0$ and find
\begin{align}
\left(\bBox-2 \kappa \right) h^{\rm (TT)}_{\mu\nu} = -\frac{2}{M_{\rm Pl}^2} \, T^{\rm (TT)}_{\mu\nu}, &&
\left(\bBox+4\kappa \right) \psi = \frac{T}{6M_{\rm Pl}^2} \;.\nonumber
\end{align}
Here
\begin{align}
T^{\rm (TT)}_{\mu\nu}=T_{\mu\nu}+\frac13 \left[\bnabla_\mu
\bnabla_\nu-\bg_{\mu\nu} \, (\bBox+3\kappa) \right]
  \left(\frac{1}{\bBox+4\kappa}\Big|_s T\right)\;, \nonumber
\end{align}
and $\frac{1}{\bBox-\mu^2}\big|_{s, t}$ is the inverse of the operator
$\bBox-\mu^2$ acting on scalars ($s$) and tensors ($t$), respectively.
The corresponding Green's functions are assumed to respect Lorentz
invariance, and satisfy
$\frac{1}{\bBox-\mu^2}\big|_{s} 1=-\frac{1}{\mu^2}$, for
$\mu^2 \neq 0$. For $\mu^2=0$ and on a Minkowski background, we will
also need the relation
$\partial_{\mu} \partial_{\nu} \frac{1}{\bBox}\big |_s 1=\frac14
\eta_{\mu\nu}$, as motivated by Lorentz invariance.

It follows that
$ {\cal A} \big|_{m^2=0} = \bar {\cal A}_{2} / M_{\rm Pl}^2$, where
\begin{multline}\label{eq:A0gen}
  \bar {\cal A}_{2} = -\int \rd^4 x \, \sqrt{-\bg} \left\{T'^{\mu\nu}_{\rm (TT)} \frac{1}{\bBox-2\kappa}\Big|_{ t} T^{\rm (TT)}_{\mu\nu} \right. \\
  \left. -\frac16 \, T' \frac{1}{\bBox+4 \kappa} \Big|_s  T  \right\} \;.
\end{multline}
For {\it localised} conserved sources, we can manipulate this
expression into a more familiar form
\begin{multline}
  \bar {\cal A}_2 = - \int \rd^4 x \, \sqrt{-\bg} \left\{T'^{\mu\nu} \frac{1}{\bBox-2\kappa}\Big|_{ t} T_{\mu\nu} \right. \\
  \left. -\frac12 T' \left[ \frac{\frac12}{\bBox-2\kappa}\Big|_s + \frac{\frac12}{\bBox+6\kappa}\Big|_s \right] T \right\} \;.\label{A0}
\end{multline}
where we have used the fact that for localised $X(x)$ and $Y(x)$, we
have
\begin{multline}
\frac{1}{\bBox-\mu^2}\Big|_{ t} \bnabla_\mu\bnabla_\nu \, X=\bnabla_\mu\bnabla_\nu \, \frac{1}{\bBox-\mu^2+8\kappa}\Big|_{ s} X\\+\frac14 \, \bg_{\mu\nu}\left[\frac{\mu^2 }{\bBox-\mu^2}\Big|_{ s} +\frac{8\kappa-\mu^2}{\bBox-\mu^2+8\kappa}\Big|_{ s}\right] X \;, \label{commute}
\end{multline}
and
$\frac{1}{\bBox-\mu^2}\Big|_{ t} \bg_{\mu\nu} Y=\bg_{\mu\nu}
\frac{1}{\bBox-\mu^2}\Big|_{ s} Y$.  Note that Eq.~\eqref{commute} was
not taken into account in \cite{adsvdvz1}, which accounts for the
subtle differences compared to our formula in Eq.~\eqref{A0} [and
Eq.~\eqref{Am}].

For $m^2 \neq 0,\, 2\kappa $,\footnote{{The partially massless
    case \cite{Deser} with $m^2 = 2\kappa$  is not considered here since it has been argued to contain an infinitely strongly coupled helcility-0 mode in violation of perturbative unitarity \cite{PM1, PM2, PM3}.}} we obtain
$A_\mu = 0$, $\psi = \kappa \, \phi$ and
\begin{align}
  \left( \bBox-2 \kappa-m^2 \right) h^{\rm (TT)}_{\mu\nu} &= -\frac{2}{M_{\rm Pl}^2} \, T^{\rm (TT)}_{\mu\nu}\;, \nonumber \\
  \left(\bBox+4\kappa\right)\phi& = \frac{T}{3 \left(2\kappa-m^2\right) M_{\rm Pl}^2}  \nonumber \;.
\end{align}
It follows that
${\cal A}_{m^2 \neq 0, 2\kappa} = {\cal A}_2(m^2) / M_{\rm Pl}^2$,
where
\begin{multline}\label{Amgen}
  {\cal A}_2(m^2)=- \int \rd^4 x \, \sqrt{-\bg} \Bigg\{T'^{\mu\nu}_{\rm (TT)} \frac{1}{\bBox-2\kappa-m^2}\Big|_{ t} T^{\rm (TT)}_{\mu\nu} \\
   - \frac{\kappa}{3(2\kappa-m^2)} \, T'  \frac{1}{\bBox+4\kappa}   \Big|_sT \Bigg\} \;. 
\end{multline}
This formula holds in all cases, although for \textit{two}
non-localised, but constant, vacuum energy sources care must be taken
to take the $\kappa=0$ limit only {\it after} identifying
$ \frac{1}{\bBox+4\kappa}$ with $ \frac{1}{4\kappa}$.

For two {\it localised} conserved sources, we can once again
manipulate this formula into the following form,
\begin{multline}\label{Am}
  {\cal A}_2(m^2)=- \int \rd^4 x \, \sqrt{-\bg} \left\{T'^{\mu\nu} \frac{1}{\bBox-2\kappa-m^2}\Big|_{ t} T_{\mu\nu} \right. \\
  \left. -\frac12 \, T' \left[ \frac{1/2}{\bBox-2\kappa-m^2} + \frac{\kappa-m^2/6}{\kappa-m^2/2} \,\frac{1/2}{\bBox+6\kappa-m^2} \right]_s T \right\}\;. 
\end{multline}
Notice the absence of the vDVZ discontinuity \cite{vdv,z} as $m^2 \to 0$ for $\kappa \neq 0$ \cite{adsvdvz1,adsvdvz2}.

\end{document}